\documentclass{ws-procs975x65}

\newcommand{\br}{\bm{r}}

\begin{document}

\title{DEFORMED RELATIVISTIC HARTREE-BOGOLIUBOV MODEL FOR EXOTIC NUCLEI}

\author{S. G. Zhou$^{*}$}

\address{Institute of Theoretical Physics,
Chinese Academy of Sciences, Beijing 100080, China\\
Center of Theoretical Nuclear Physics, National Laboratory
              of Heavy Ion Accelerator, \\Lanzhou 730000, China\\
 $^*$E-mail: sgzhou@itp.ac.cn}

\author{J. Meng}

\address{School of Physics, Peking University, Beijing 100871, China \\
Institute of Theoretical Physics,
Chinese Academy of Sciences, Beijing 100080, China\\
Center of Theoretical Nuclear Physics, National Laboratory
              of Heavy Ion Accelerator, \\Lanzhou 730000, China\\
}

\author{P. Ring}

\address{
Physikdepartment, Technische Universit\"at M\"unchen,
              85748 Garching, Germany \\
}

\begin{abstract}
A deformed relativistic Hartree-Bogoliubov (DRHB) model is developed
aiming at a proper description of exotic nuclei, particularly
deformed ones with large spatial extension. In order to give an
adequate description of both the contribution of the continuum and
the large spatial distribution in exotic nuclei, the DRHB equations
are solved in a Woods-Saxon basis in which the radial wave functions
have proper asymptotic behaviors at large distance from the nuclear
center which is crucial for the formation of halo. The formalism and
the numerical procedure of the DRHB model in a Woods-Saxon basis are
briefly presented.
\end{abstract}


\bodymatter

\section{\label{sec:intro}Introduction}

One of the exotic phenomena observed in nuclei close to drip lines
is the halo in which the extremely weakly binding property leads to
many new features, e.g., the coupling between the bound states and
the continuum due to pairing correlations and the large spacial
density distribution. In order to give an adequate theoretical
description of the halo phenomenon, the asymptotic behavior of
nuclear densities at large $r$ must be considered properly and the
discrete bound states, the continuum and the coupling between them
must be treated self consistently. This could be achieved by solving
the non-relativistic Hartree-Fock-Bogoliubov (HFB)~\cite{Bulgac80,
Dobaczewski84, Dobaczewski96} or relativistic Hartree-Bogoliubov
(RHB)~\cite{Meng96, Poeschl97, Meng98} equations in coordinate space
which can fully take into account the mean-field effects of the
coupling to the continuum.

In Refs.~\cite{Dobaczewski84, Dobaczewski96, Meng96, Poeschl97,
Meng98}, the spherical symmetry is assumed. Since most of the known
nuclei are deformed, whether or not there exist deformed halos and
what new features are expected in deformed exotic nuclei are very
interesting questions~\cite{Li96, Misu97, Guo03, Meng03, Nunes05,
Pei06} which could be answered by the deformed counterparts of the
HFB or RHB models. Nevertheless for deformed nuclei, to solve the
HFB or RHB equations in coordinate space becomes much more
sophisticated and numerically very time consuming. Although many
efforts have been made to develop non-relativistic HFB models either
in (discretized) coordinate space or in a basis with improved
asymptotic behavior~\cite{Terasaki96, Terasaki97, Yamagami01,
Stoitsov00, Stoitsov03, Oberacker03a, Oberacker03b, Tajima04}, the
deformed relativistic Hartree-Bogoliubov model has been developed
only in conventional harmonic oscillator basis~\cite{Vretenar99,
Lalazissis99, Niksic02, Niksic04}.

The deformed relativistic Hartree equations have been solved in a
basis in which the basis wave functions are calculated from solving
the Dirac equation with spherical Hartree potentials~\cite{Zhang88,
Zhang91}. In Ref.~\cite{Zhou03}, the Woods-Saxon basis was proposed
as a reconciler between the harmonic oscillator basis and the
coordinate space. The Woods-Saxon wave functions have more realistic
asymptotic behavior at large $r$. One can also use a box boundary
condition to discretize the continuum. It has been shown that to
solve the spherical relativistic Hartree equations in a Woods-Saxon
basis is almost equivalent to to do it in coordinate
space~\cite{Zhou03}. The Woods-Saxon basis may be used in more
complicated situations, e.g., for the description of exotic nuclei
where both deformation and pairing have to be taken into
account~\cite{Zhou06}. In this proceeding paper we will present
briefly the formalism, numerical procedure and some results of the
deformed relativistic Hartree Bogoliubov model in a Woods-Saxon
basis. Throughout the paper, this model will be labeled as the
DRHBWS model.

The paper is organized as follows. In Sec.~\ref{sec:formalism}, we
give the formalism of the DRHBWS model. The numerical procedure and
some preliminary results are presented in Sec.~\ref{sec:results}. A
summary is given in Sec.~\ref{sec:summary}.

\section{\label{sec:formalism}Formalism}

The starting point of the relativistic Hartree theory is a
Lagrangian density where nucleons are described as Dirac spinors
which interact via the exchanges of several mesons ($\sigma$,
$\omega$, and $\rho$) and the photon~\cite{Serot86, Reinhard89,
Ring96, Vretenar05, Meng06},
\begin{eqnarray}
\displaystyle
 {\cal L}
   & = &
     \bar\psi_i \left( i\rlap{/}\partial -M \right) \psi_i
    + \frac{1}{2} \partial_\mu \sigma \partial^\mu \sigma
    - U(\sigma)
    - g_{\sigma} \bar\psi_i \sigma \psi_i
   \nonumber \\
   &   & \mbox{}
    - \frac{1}{4} \Omega_{\mu\nu} \Omega^{\mu\nu}
    + \frac{1}{2} m_\omega^2 \omega_\mu \omega^\mu
    - g_{\omega} \bar\psi_i \rlap{/}\omega \psi_i
   \nonumber \\
   &   & \mbox{}
    - \frac{1}{4} \vec{R}_{\mu\nu} \vec{R}^{\mu\nu}
    + \frac{1}{2} m_{\rho}^{2} \vec{\rho}_\mu \vec{\rho}^\mu
    - g_{\rho} \bar\psi_i \rlap{/} \vec{\rho} \vec{\tau} \psi_i
   \nonumber \\
   &   &\mbox{}
    - \frac{1}{4} F_{\mu\nu} F^{\mu\nu}
    - e \bar\psi_i \frac{1-\tau_3}{2}\rlap{/}A \psi_i ,
\label{eq:Lagrangian}
\end{eqnarray}
with the summation convention used, $\rlap{/} x \equiv \gamma^\mu
x_\mu = \gamma_\mu x^\mu$, $M$ the nucleon mass, and $m_\sigma$,
$g_\sigma$, $m_\omega$, $g_\omega$, $m_\rho$, $g_\rho$ masses and
coupling constants of the respective mesons. The nonlinear
self-coupling for the scalar mesons is crucial for a satisfactory
description of the surface properties~\cite{Boguta77},
\begin{equation}
   U(\sigma) = \displaystyle\frac{1}{2} m^2_\sigma \sigma^2
              +\displaystyle\frac{g_2}{3}\sigma^3
              + \displaystyle\frac{g_3}{4}\sigma^4 ,
\end{equation}
and field tensors for the vector mesons and the photon fields are
defined as
\begin{eqnarray}
 \left\{
  \begin{array}{rcl}
   \Omega_{\mu\nu}  & = & \partial_\mu\omega_\nu
                         -\partial_\nu\omega_\mu, \\
   \vec{R}_{\mu\nu} & = & \partial_\mu\vec{\rho}_\nu
                         -\partial_\nu\vec{\rho}_\mu
                         -g_{\rho} (\vec{\rho}_\mu
                                    \times \vec{\rho}_\nu ), \\
   F_{\mu\nu}       & = & \partial_\mu {A}_\nu
                         - \partial_\nu {A}_\mu.
  \end{array}
 \right.
 \label{eq:tensors}
\end{eqnarray}
For the ground state of nuclei with time reversal symmetry, the
nucleon spinors are the eigenvectors of the stationary Dirac
equation,
\begin{equation}
  \left[ \bm{\alpha} \cdot \bm{p} + V(\bm{r}) + \beta (M + S(\bm{r}))
  \right] \psi_i(\bm{r}) = \epsilon_i \psi_i(\bm{r}) ,
\label{eq:Dirac0}
\end{equation}
and equations of motion for the mesons and the photon are
\begin{eqnarray}
 \left\{
   \begin{array}{rcl}
    \left( -\Delta + \partial_\sigma U(\sigma) \right )\sigma(\bm{r})
      & = & -g_\sigma \rho_s(\bm{r}) , \\
    \left( -\Delta + m_\omega^2 \right )             \omega^0(\bm{r})
      & = &  g_\omega \rho_v(\bm{r}) , \\
    \left( -\Delta + m_\rho^2 \right)                  \rho^0(\bm{r})
      & = &  g_\rho   \rho_3(\bm{r}) , \\
    -\Delta                                               A^0(\bm{r})
      & = &  e        \rho_p(\bm{r}) ,
   \end{array}
 \right.
 \label{eq:mesonmotion}
\end{eqnarray}
where $\omega^0$ and $A^0$ are time-like components of the vector
$\omega$ and the photon fields and $\rho^0$ the 3-component of the
time-like component of the iso-vector vector $\rho$ meson.
Equations~(\ref{eq:Dirac0}) and (\ref{eq:mesonmotion}) are coupled
by the vector and scalar potentials,
\begin{eqnarray}
 \left\{
   \begin{array}{lll}
     V(\bm{r}) & = & g_\omega \omega^0(\bm{r})
                    +g_\rho \tau_3 \rho^0(\bm{r})
                    +e \displaystyle\frac{1-\tau_3}{2} A^0(\bm{r}) , \\
     S(\bm{r}) & = & g_\sigma \sigma(\bm{r}), \\
   \end{array}
 \right.
 \label{eq:vaspot}
\end{eqnarray}
and various densities
\begin{eqnarray}
 \left\{
  \begin{array}{rcl}
   \rho_s(\bm{r})
   & = &
    \sum_{i=1}^A \bar\psi_i(\bm{r}) \psi_i(\bm{r}) ,\\
   \rho_v(\bm{r})
   & = &
    \sum_{i=1}^A \psi_i^\dagger(\bm{r}) \psi_i(\bm{r}) ,\\
   \rho_3(\bm{r})
   & = &
    \sum_{i=1}^A \psi_i^\dagger(\bm{r}) \tau_3 \psi_i(\bm{r}) ,\\
   \rho_c(\bm{r})
   & = &
    \sum_{i=1}^A \psi_i^\dagger(\bm{r})
                 \displaystyle\frac{1-\tau_3}{2}\psi_i(\bm{r}) .
  \end{array}
 \right.
 \label{eq:mesonsource}
\end{eqnarray}

The pairing correlation is included via the Bogoliubov
transformation and the relativistic Hartree-Bogoliubov equation
reads~\cite{Kucharek91},
\begin{eqnarray}
 \sum_{\sigma'p'} \int d^3 \br'
 \left(
  \begin{array}{cc}
   h_D(\br\sigma p,\br\sigma'p') - \lambda &
   \Delta(\br\sigma p,\br'\sigma' p') \\
  -\Delta^*(\br\sigma p,\br'\sigma' p')
   & -h_D(\br\sigma p,\br\sigma'p') + \lambda \\
  \end{array}
 \right)
 \left(
  { U_{k}(\br'\sigma' p') \atop V_{k}(\br'\sigma' p') }
 \right)
 &  & \nonumber \\
 =
 E_{k}
  \left(
   { U_{k}(\br\sigma p) \atop V_{k}(\br\sigma p) }
  \right)
 ,
 &  &
 \label{eq:RHB0}
\end{eqnarray}
where $E_{k}$ is the quasiparticle energy, $h_D$ the Dirac
Hamiltonian in (\ref{eq:Dirac0}), and $\lambda$ the Fermi energy.
Here $p=1, 2$ (or $\pm$) is used to represent the
particle-antiparticle degree of freedom.

If in the pp channel, we use a zero range density dependent force,
\begin{equation}
 V_{p_1p_2p'_1p'_2}(\br_1,\br_2;\sigma_1\sigma_2\sigma'_1\sigma'_2)
 = \frac{1}{4} V_0 \delta( \mathbf{r}_1 - \mathbf{r}_2 )
   \left[ 1 - 4 \vec{\sigma}_{11'} \cdot \vec{\sigma}_{22'} \right]
   \left[ \mathbf{I}^p_{11'} \cdot \mathbf{I}^p_{22'} \right]
 ,
 \label{eq:pairing_force}
\end{equation}
for axially deformed nuclei with spacial reflection symmetry, we may
expand the potentials, $S(\bm{r})$, $V(\bm{r})$, and $\Delta(\bm
r)$, and various densities in terms of the Legendre polynomials,
\begin{equation}
 f(\bm{r})   = \sum_\lambda f_\lambda({r}) P_\lambda(\cos\theta),\
 f_\lambda(r) = \frac{2\lambda+1}{2}
                \int d\cos\theta f(\bm{r})  P_\lambda(\cos\theta),\
 \lambda = 0,2,4,\cdots.
 \label{eq:expansion}
\end{equation}
The quasi particle wave function is expanded in the Woods-Saxon
basis $\left\{ \epsilon_{i\kappa m}, \varphi_{n\kappa
m}(\bm{r}\sigma p) \right\}$ as,
\begin{eqnarray}
   U_{k} (\br\sigma p)
 & = & \displaystyle
  \sum_{i\kappa}
  \left(
  { u^{(m)}_{k,(i\kappa)} \varphi_{i\kappa m}(\br\sigma p)
     \atop
     u^{(\bar m)}_{k,(\widetilde{i\kappa})} \tilde\varphi_{i\kappa m}(\br\sigma p)
   }
  \right)
  ,\nonumber \\
   V_{k} (\br\sigma p)
 & = & \displaystyle
  \sum_{i\kappa}
  \left(
   {
     v^{(m)}_{k,({i\kappa})} \varphi_{i\kappa m}(\br\sigma p)
     \atop
     v^{(\bar m)}_{k,(\widetilde{i\kappa})} \tilde\varphi_{i\kappa m}(\br\sigma p)
   }
  \right)
 ,
 \label{eq:UVexpansion}
\end{eqnarray}
where the single particle energy $\epsilon_{i\kappa m}$ and wave
function $\varphi_{i\kappa m}(\bm{r}\sigma p)$ are obtained from
solving the spherical Dirac equation with Woods-Saxon-like
potentials~\cite{Koepf91},
\begin{equation}
 \varphi_{i\kappa m}(\bm{r}\sigma) =
   \frac{1}{r}
   \left(
     \begin{array}{c}
       i G_{i\kappa}(r) Y^l _{jm} (\theta,\phi,\sigma)
       \\
       - F_{i\kappa}(r) Y^{\tilde l}_{jm}(\theta,\phi,\sigma)
     \end{array}
   \right) ,
   \ \ j = l\pm\frac{1}{2},
 \label{eq:SRHspinor}
\end{equation}
with $G_\alpha^{\kappa}(r) / r$ and $F_\alpha^{\kappa}(r) / r$ the
radial wave functions for the upper and lower components and $Y^l
_{jm}(\theta,\phi,\sigma)$ the spin spherical harmonics where
$\kappa = (-1)^{j+l+1/2} (j+1/2)$ and $\tilde l = l +
(-1)^{j+l-1/2}$. The states both in the Fermi sea and in the Dirac
sea should be included in the basis for the completeness. In the
Woods-Saxon basis, for each $m$-block the RHB equation
(\ref{eq:RHB0}) turns out to be,
\begin{equation}
 \left( \begin{array}{cc}
  {\cal A} & {\cal B} \\
  {\cal C} & {\cal D} \\
 \end{array} \right)
 \left(
  { {\cal U}
    \atop
    {\cal V}
  }
 \right)
 = E
 \left(
  { {\cal U}
    \atop
    {\cal V}
  }
 \right),
\end{equation}
where
\begin{equation}
 {\cal U} = \left(u^{(m)}_{k,(i\kappa)}\right),\
 {\cal V} = \left(v^{(m)}_{k,(\widetilde{i\kappa})}\right),
\end{equation}
\begin{equation}
 {\cal A} = \left(A^{(m)}_{(i\kappa)(i'\kappa')}\right),\
 {\cal D} =
  \left(-A^{(m)}_{(\widetilde{i\kappa})(\widetilde{i'\kappa'})}\right),
\end{equation}
\begin{equation}
 {\cal B} =
  \left(\Delta^{(m)}_{(i\kappa)(\widetilde{i'\kappa'})}\right),\
 {\cal C} =
  \left(-\Delta^{(m)}_{(\widetilde{i\kappa})(i'\kappa')}
 =\Delta^{(m)}_{(i'\kappa')(\widetilde{i\kappa})} \right).
 \label{eq:pairing_matrix}
\end{equation}
The derivation of the matrix elements will be given in a detailed
paper~\cite{Zhou07}.

\section{\label{sec:results}Numerical procedure and results}

\begin{figure}
\begin{center}
\includegraphics[width=7cm]{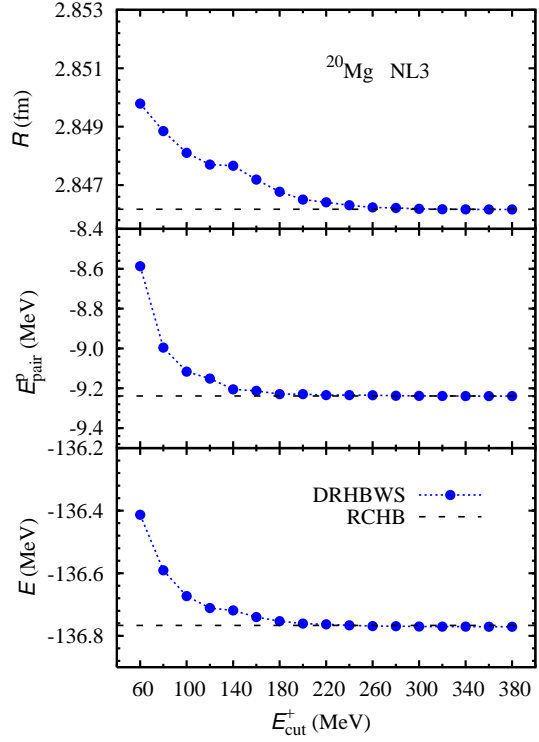}
\end{center}
\caption{\label{fig:mg20_conv} The total binding energy $E$ (the
lower panel), the proton pairing energy $E^\mathrm{p}_\mathrm{pair}$
(the middle panel), and the rms radius $R$ (the upper panel) of
$^{20}$Mg versus the cutoff energy $E^+_\mathrm{cut}$ in the
Woods-Saxon basis for the DRHBWS model (solid circles). The
spherical RCHB results (dashed lines) are also included for
comparison. }
\end{figure}

There are several parameters which have to be introduced for
numerical reasons, e.g., the mesh size $\Delta r$, the box size
$R_\mathrm{max}$, and the cut off parameter on $\lambda$ in the
expansion (\ref{eq:expansion}), $\lambda_\mathrm{max}$. Instead of a
cut off on the radial quantum number $n$ in the expansion
(\ref{eq:UVexpansion}), an energy cutoff $E^+_\mathrm{cut}$ is
introduced for positive energy states in the Woods-Saxon basis and
in each $\kappa$-block, the number of negative energy states in the
Dirac sea is the same as that of positive energy states in the Fermi
sea. We have investigated the dependence of our results on some of
these parameters in spherical and deformed relativistic Hartree
models~\cite{Zhou03, Zhou06}. It's found that a box of the size
$R_\mathrm{max} = 4r_0 A^{1/3}$ with $r_0 = 1.2$ fm, a step size
$\Delta r = 0.1$ fm, $\lambda_\mathrm{max} = 4$, and
$E^+_\mathrm{cut} = 100$ MeV give relative deviations of the binding
energy, the rms radius, and the quadrupole moment from the standard
ones smaller than 0.1 \% for light nuclei. In order to reduce the
computational time, a small cutoff $\lambda_\mathrm{max} = 3$ is
used which would not introduce sizable errors. The parameter set
NL3~\cite{Lalazissis97} is used for the Lagrangian density.

There are two parameters ($\rho_0$ and $V_0$) in the
phenomenological pairing force (\ref{eq:pairing_force}). Since we
are using a zero range force, a cutoff
$E^\mathrm{q.p.}_\mathrm{cut}$ must be introduced to define the
pairing window. We take the empirical value 0.152 fm$^{-3}$ for the
saturation density $\rho_0$. The pairing strength $V_0 = 374$ MeV
fm$^3$ and the cutoff $E^\mathrm{q.p.}_\mathrm{cut} = 60$ MeV
reproduce the proton pairing energy in the spherical nucleus
$^{20}$Mg from the spherical relativistic Hartree Bogoliubov theory
in a harmonic oscillator basis in which the Gogny force is used in
the pp channel.

As the first application of the DRHBWS method, we study a spherical
nucleus $^{20}$Mg so that a comparison can be made between the
DRHBWS and the spherical RCHB~\cite{Meng96, Meng98} results. The
ground state properties of $^{20}$Mg are calculated with the DRHBWS
and the RCHB codes. The total binding energy $E$, the proton pairing
energy $E^\mathrm{p}_\mathrm{pair}$, and the rms radius $R$ of
$^{20}$Mg are plotted versus $E^+_\text{cut}$ in the lowest, middle
and top panels of Fig.~\ref{fig:mg20_conv}, respectively. The
spherical RCHB results, regarded as exact ones, are shown as
horizontal dashed lines. When the basis size increases, $E$,
$E^\mathrm{p}_\mathrm{pair}$, and $R$ all converge to the
corresponding exact values. In practical calculations, one can
choose $E^+_\text{cut}$ according to the balance between the desired
accuracy and the computational cost. For light nuclei, one can
safely use $E^+_\text{cut}$ = 100 MeV which results in accuracies in
the total binding energy and the proton pairing energy of about a
hundred keV and in the rms radius of around 0.002 fm.

We are investigating some light deformed nuclei close to the neutron
drip line by using the DRHBWS model. The results will be presented
elsewhere~\cite{Zhou07}.

\section{\label{sec:summary}Summary}

In order to give a proper description of exotic nuclei, particularly
deformed ones with large spatial extension, the deformed
relativistic Hartree-Bogoliubov (DRHB) equations are solved in a
Woods-Saxon basis in which the radial wave functions have proper
asymptotic behaviors at large distance from the nuclear center which
is crucial for the formation of halo.

In this contribution, the formalism and the numerical procedure of
the DRHB model in a Woods-Saxon basis are briefly presented. Some
preliminary results, namely, the results for a spherical nucleus
$^{20}$Mg is also given and compared with the spherical relativistic
Hartree-Bogoliubov model.

\section*{Acknowledgments}
This work was supported by the National Natural Science Foundation
of China under Grant Nos. 10435010, 10475003 and 10575036, the Major
State Basic Research Development Program of China under contract No.
2007CB815000, the Knowledge Innovation Project of Chinese Academy of
Sciences under contract Nos. KJCX3-SYW-N02 and KJCX2-SW-N17, and
Asia Link Programme CN/Asia-Link 008 (94 791). The computation of
this work was performed on the HP-SC45 Sigma-X parallel computer of
ITP and ICTS and supported by Supercomputing Center, CNIC, Chinese
Academy of Sciences.

\end{document}